\documentclass[sigconf,9pt,usenames,dvipsnames,table]{acmart}

\AtBeginDocument{%
  \providecommand\BibTeX{{%
    \normalfont B\kern-0.5em{\scshape i\kern-0.25em b}\kern-0.8em\TeX}}}

\setcopyright{rightsretained}
\copyrightyear{2024}
\acmYear{2024}
\acmDOI{10.48550/arXiv.2506.23866}
\acmConference[LOCO '24]{LOCO '24: 1st International Workshop on Low Carbon
Computing}{December 03, 2024}{Glasgow, UK/Online}
\acmBooktitle{LOCO '24: 1st International Workshop on Low Carbon 
Computing, December 03, 2024}
\acmISBN{}

\usepackage{balance}
\usepackage{verbatim}
\usepackage[textsize=footnotesize, textwidth=15mm]{todonotes}
\presetkeys{todonotes}{fancyline}{}
\usepackage[english=american,babel=false]{csquotes}
\usepackage{nth}

\usepackage{acronym}
\acrodef{EU}{European Union}

\usepackage{microtype}
\usepackage{float}
\usepackage{placeins}
\usepackage{enumitem}
\usepackage{subcaption}

\usepackage[]{multibib}
\newcites{Tools}{Tools References}

\definecolor{bw1}{RGB}{0,114,178}
\definecolor{bw2}{RGB}{230,159,0}
\definecolor{bw3}{RGB}{86,180,233}
\definecolor{bw4}{RGB}{0,158,115}
\definecolor{bw5}{RGB}{240,228,66}

\usepackage{xcolor}

\makeatletter
\newcommand*\bigcdot{\mathpalette\bigcdot@{2}}
\newcommand*\bigcdot@[2]{\mathbin{\vcenter{\hbox{\scalebox{#2}{$\m@th#1\bullet$}}}}}
\makeatother

\begin{document}

\title[Exploring Privacy and Security
  as Drivers for Environmental
  Sustainability in Cloud-Based
  Office Solutions]{Exploring Privacy and Security
  as Drivers for Environmental
  Sustainability in Cloud-Based
  Office Solutions}

\author{Jason Kayembe}
\email{jason.kayembe@ulb.be}
\orcid{0009-0003-1279-2666}
\affiliation{%
  \institution{Universit\'e Libre de Bruxelles}
  \city{Brussels}
  \country{Belgium}
  \postcode{1050}
}

\author{Iness Ben Guirat}
\email{iness.ben.guirat@ulb.be}
\orcid{0000-0002-8766-594X}
\affiliation{%
  \institution{Universit\'e Libre de Bruxelles}
  \city{Brussels}
  \country{Belgium}
  \postcode{1050}
}

\author{Jan Tobias M\"uhlberg}
\email{jan.tobias.muehlberg@ulb.be}
\orcid{0000-0001-5035-0576}
\affiliation{%
  \institution{Universit\'e Libre de Bruxelles}
  \city{Brussels}
  \country{Belgium}
  \postcode{1050}
}

\begin{abstract}
  In this paper, we explore the intersection of privacy, security, and environmental sustainability in cloud-based office solutions, focusing on quantifying user- and network-side energy use and associated carbon emissions. We hypothesise that privacy-focused services are typically more energy-efficient than those funded through data collection and advertising. To evaluate this, we propose a framework that systematically measures environmental costs based on energy usage and network data traffic during well-defined, automated usage scenarios. To test our hypothesis, we first analyse how underlying architectures and business models, such as monetisation through personalised advertising, contribute to the environmental footprint of these services. We then explore existing methodologies and tools for software environmental impact assessment.

We apply our framework to three mainstream email services selected to reflect different privacy policies, from ad-supported tracking-intensive models to privacy-focused designs: Microsoft Outlook, Google Mail (Gmail), and Proton Mail. We extend this comparison to a self-hosted email solution, evaluated with and without end-to-end encryption. We show that the self-hosted solution, even with 14\% of device energy and 15\% of emissions overheads from PGP encryption, remains the most energy-efficient, saving up to 33\% of emissions per session compared to Gmail. Among commercial providers, Proton Mail is the most efficient, saving up to 0.1 gCO\(_2\)e per session compared to Outlook, whose emissions can be further reduced by 2\% through ad-blocking.

At scale, these differences become significant: switching all active Gmail users to a self-hosted solution (without PGP) could save up to 11.9 ktCO\(_2\)e annually — equivalent to the emissions of approximately 9,000 round-trip flights Paris - New York. Similarly, moving all Outlook users to Proton Mail could save up to 2.2 ktCO\(_2\)e annually, while enabling ad-blocking for all of them could prevent an estimated 448 tCO\(_2\)e per year. Additionally, we observe that the network I/O for Outlook and Gmail is more than twice that of Proton Mail, reflecting the higher data transfer linked to advertising and tracking. These results support our hypothesis that privacy-preserving design choices not only enhance data protection but also reduce environmental impact — suggesting that privacy and sustainability can be mutually reinforcing in the design of online services.
\end{abstract}

\keywords{privacy, security, webmail, advertising, environmental impacts}

\maketitle
\renewcommand{\shortauthors}{Kayembe et al.}

\section{Introduction}
Concerns over the environmental impact of human activity, alongside with
growing privacy and security issues, motivate this work to investigate if
these challenges and potential mitigations are interconnected. As an
example, the European Data Protection Supervisor (EDPS) found that the
European Commission’s use of Microsoft 365 violates EU data protection
regulations due to unclear data collection, purpose, and processing
location~\cite{EuropeanCommissionsUse2024}. Following our intuition, a
service that only collects and processes the data that is strictly
necessary for delivering that service would, overall, have a smaller
environmental footprint than a service that collects and shares personal
data extensively and unnecessarily. We thus investigate if privacy and
security are drivers for energy-efficient solutions.

Although the example of the EDPS prematurely warns us about the difficulty
to dress the complete picture, this paper investigates the intersection of
privacy, security, and the environmental consequences of ICT services. We
place a particular focus on web-based email services as an example of
commonly used office solutions. We explore how to define and evaluate these
services, examine the architectures and business models they rely on, and
envision a framework that can measure the energy consumption associated
with specific cloud services. Our research hypothesis is that \emph{online
  services that emphasise privacy and security are typically
  \enquote{greener}}, and we use web mailers as a case study to test this
hypothesis.

\paragraph{Environmental Impacts of Webmail Services.}

Sending an email might seem like a trivial task; however, the scale at which large email services operate makes their environmental impact significant. According to Statista, an estimated 333 billion emails were sent and received daily worldwide in 2022~\cite{statistaemails2024}. More generally, email services are part of a broader ecosystem of cloud-based office tools---including document editing platforms, cloud storage, calendar systems, and collaboration suites---that all rely on internet infrastructure. The environmental footprint of these services arises across three main components: user-side, network-side, and server-side emissions. This is illustrated in~\cite{parssinen_environmental_2018} where P\"arssinen et al. study the environmental cost of online advertising and take inventory of common internet network devices to derive the overall environmental impact of the industry.

Depending on their business model, services may often deploy additional technologies that go beyond their core functionalities, leading to increased energy consumption, such as client-side ad rendering and user tracking scripts, and server-side operations like personalised model training. Additionally, the sending and rendering of advertisements monopolises resources. P\"arssinen et al.~\cite{parssinen_environmental_2018} describe the mechanisms behind this: When a user visits a web page, this often initiates thousands of connections to data centres. Some of these data centres keep the connections open to deliver advertisements as long as the user remains on the page, thereby requiring network, data centre, and user device resources. This is further supported by Pearce et al.~\cite{pearce2020energy}, who quantify the energy savings achievable through the use of ad-blockers, showing that page load times can decrease by up to 28\,\%. This represents potential energy savings of 13.5 billion kWh per year for internet users worldwide.

In contrast, privacy-focused services tend to limit processing to what is strictly necessary, reducing data transfer and avoiding resource-intensive advertising mechanisms. We propose a framework allowing to highlight the tangible environmental benefits of such privacy-preserving approaches. By systematically measuring user-side energy consumption and network data traffic during defined usage scenarios, our method enables comparisons between services with similar functionalities but differing business models. While our current application focuses on webmail, the framework is designed to generalise across web-based services. Adapting it only requires redefining adequate functional units and implementing their execution via automation scripts.

\paragraph{This Paper.}

Business models centred on data collection and (personalised)
advertising raise not only privacy and security concerns but also
exacerbate environmental costs. Thus, validating the relationship between
business models and energy consumption can guide the design of more secure,
privacy-friendly, and also more sustainable online services.  This paper
compares the environmental burdens of three email service providers:
Microsoft Outlook and Google's Gmail, both of which rely heavily on
advertising and tracking technologies, and Proton Mail, which operates on a
privacy-preserving and end-to-end secure freemium model without the use of
advertisements or trackers. To understand the environmental cost of
processing advertisements on the user side, we evaluate the effect of using
a DNS-based ad-blocker on the energy consumption of the email services.
Additionally, we set up our own email server with two objectives in mind:
Comparing the environmental impact of self-hosted email services to those
of the major providers, and assessing the impact of PGP encryption on
energy consumption. Lastly, we validate our framework against the study by
Pearce et al.~\cite{pearce2020energy}, which evaluates the energy savings
from reduced page loading times due to ad-blockers on advertisement-heavy
websites. Comparing their results helps us confirm our tool's capability to
measure the energy impact of ad-blockers. We provide a comprehensive
evaluation of the environmental impact of email services, and explore
the potential of privacy and security as drivers for environmental
sustainability in cloud-based office solutions. We make our system
configurations and research data available at~\citeTools{artifact-jason}.

\paragraph{Insights.}

The main insight from our experimental results is that online services prioritising privacy and security such as Proton Mail and self-hosted solutions demonstrate lower user-side energy consumption and network data transfer compared to mainstream, ad-supported providers like Gmail and Outlook. Advertising and tracking mechanisms significantly increase both energy use and data traffic, as evidenced by the measurable reductions achieved through ad-blocking. While enabling strong security features like PGP encryption does incur a modest energy and data overhead, the overall environmental impact remains lower than that of ad-driven services. These findings support the hypothesis that privacy- and security-focused online services tend to be more environmentally sustainable, and highlight the potential for further efficiency improvements in mainstream solutions.


\section{Related Work}

Our work builds on prior research across three main areas: the environmental footprint of online advertising, the methodological foundations of environmental impact assessment in ICT, and the relationship between privacy, sustainability, and digital sovereignty.

\paragraph{Assessing the environmental footprint of advertising.}
Few studies have addressed the energy and emissions overhead introduced by online advertising and tracking. P\"arssinen et al.~\cite{parssinen_environmental_2018} proposed a layered framework that combines bottom-up inventory and impact assessment of the whole internet system, with top-down allocation of the emissions based on types of traffic protocols, offering a way to attribute system-level emissions to specific categories of service---such as advertisement. Pesari et al.~\cite{pesari2023client} applied a client-side measurement approach to quantify the energy cost of ads and trackers on news websites, while Pearce et al.~\cite{pearce2020energy} showed that ad-blockers can reduce energy use by shortening page load times. These studies focus either on broad service categories or on narrowly defined user actions. Our contribution builds on this work by enabling the assessment of more complex user interactions, such as webmail use, through the decomposition of complex scenarios into small, well-defined functional units. This allows for benchmarking of individual operations in terms of energy, emissions, data traffic, and execution time.

\paragraph{Methodological Foundations.}
Environmental assessments in ICT generally follow either top-down or bottom-up approaches. Top-down methods rely on sector-wide data and allocation principles. For example, Malmodin et al.~\cite{malmodin2024ict,lovehagen_assessing_2023} provide global life-cycle estimates for user devices, networks, and servers, while Aslan et al.~\cite{aslanElectricityIntensityInternet2018} propose a model to allocate network emissions based on data traffic. Freitag et al.~\cite{freitag_real_2021} highlight the challenges of such estimates, pointing to uncertainties arising from differing system boundaries and underlying hypothesis. Their review can be used to select adequate estimates considering the underlying assumptions. In contrast, bottom-up approaches rely on direct measurement, using tools like the Green Metrics Tool (GMT)~\citeTools{greencoding}. Our framework builds on GMT by introducing functional units to divide and standardise complex services. This in turn, enables the isolation of specific design properties, such as privacy versus tracking mechanisms, and encryption.

\paragraph{Privacy, sustainability, and data sovereignty.}
Beyond methodology, our work contributes to ongoing discussions at the intersection of privacy and sustainability. Recent work by Fiebig et al.~\cite{fiebig2022position,fiebig202213,fiebig_heads_2023} highlights the importance of publicly governed infrastructures and digital sovereignty as a means to protect privacy and reduce environmental dependency on hyperscale cloud providers. The General Data Protection Regulation (GDPR)~\cite{Wolford_gdpr_2018}, in force since 2018, defines privacy as a fundamental right and advocates data protection “by design and by default.” By enforcing principles such as data minimisation and storage limitation, the GDPR incentivises service providers to restrict data collection and processing to what is strictly necessary. These requirements not only strengthen privacy but may also reduce energy use, suggesting a regulatory pathway toward more sustainable service design. Our findings support this connection by empirically showing that services with stronger privacy guarantees, such as Proton Mail, tend to be \emph{\enquote{greener}}. We further show that data sovereignty, when implemented through shared self-hosting by institutions, can contribute to more sustainable and autonomous digital infrastructures.

\section{Methodology}

The main contribution of this paper is a framework to evaluate the energy consumption of a web service by assessing the service's distinct, measurable functional units. The goal of our framework is twofold: first, to facilitate the comparison of similar services based on their energy consumption, and second to highlight the connection between data collection, processing, and the associated environmental costs.  We identify four key components in the evaluation: \emph{(1)} user-side, \emph{(2)} network-side, \emph{(3)} server-side, and \emph{(4)} embodied and end-of-life (EoL) emissions. In this work, we focus on the first two and the last, providing comprehensive estimations of user-side and network-side impacts.
This framework builds upon, and extends the works of Pesari et al.~\cite{pesari2023client} and Pearce et al.~\cite{pearce2020energy} that both examine the power usage of a computer visiting major websites when using ad-blockers.

\subsection{User-Side Assessment}

To evaluate the emissions released by a user device when accessing
the service, and to identify the attributable share for tracking and
advertising, our framework involves six steps:

\begin{enumerate}
  \item \textbf{Select Comparable Services}: The framework is designed to compare different implementations of online services. Different implementations may reveal differences in consumption profiles, which could be attributed to various factors distinguishing the implementations. However, in our case, the aim is not only to identify the greener solution, but also to assess the impact of advertising mechanisms on device energy consumption. Hence, we run the same service multiple times, both with and without the ad-blocker activated. Additionally, we configure the browser to allow the blocking of cookies and the restriction of certain website tracking capabilities in order to account for the impact of tracking mechanisms.
  \item \textbf{Define Functional Units}: We identify the common scenarios that users typically engage in when using the service such as logging into the account, reading and sending an email, replying to, and logging out.
  \item \textbf{Automate Scenarios}: Automate the identified scenarios for repeated execution without manual intervention.
  \item \textbf{Monitor Scenarios}: This step involves executing the scripts developed in the previous step and monitoring (i) machine power or energy consumption (in Watts or Joules) and (ii) data traffic on the network (in Bytes). This is achieved using metric tool APIs relying on modern computers' embedded sensors to track their components power consumption.
  \item \textbf{Convert Energy to Greenhouse Gas (GHG) Emissions}: This is typically done using an average grid energy mix and its carbon intensity. We use a world average of 445 gCO\(_2\)e/kWh~\citeTools{electricitymap}. We also convert energy from joules to kilowatt-hours using the factor \( 2.7778 \times 10^{-7} \) kWh/J. The resulting emission factor is:
  \begin{equation}
    \begin{aligned}
      C_{elec} = 445 \times 2.7778 \times 10^{-7} = 1.24 \times 10^{-4} \; [\text{gCO}_2\text{e}/\text{J}]
    \end{aligned}
  \end{equation}
  Given a measured energy consumption \( E^{user}_{f.u.} \) in joules for a functional unit, the use-phase CO\(_2\)-equivalent emissions are calculated as:
  \begin{equation}
    \begin{aligned}
      \Delta U^{user-side}_{f.u.} = \Delta E^{user-side}_{f.u.} \times C_{elec}
    \end{aligned}
  \end{equation}
  \item \textbf{Assess User-Side Operational Emissions}: Perform tasks 1 to 6 for both services and compute the differences in their user-side GHG emissions. Additionally, compute the differences in the network data traffic they generated and the difference in the time taken to complete the functional units.
\end{enumerate}

\subsection{Network-Side Assessment} \label{network}

As discussed in Freitag and al. work~\cite{freitag_real_2021}, there is an ongoing debate regarding the validity of allocating network emissions based on data volume. Networks are designed to handle peak activity, meaning a significant portion of their emissions remains constant regardless of data traffic. Thus, network emissions are not entirely elastic with respect to data volume. However, increasing data traffic eventually necessitates infrastructure expansion, leading to higher future emissions.
We acknowledge the limitations of this approach but we still choose to include estimates of these emissions in our framework to highlight the importance of considering the network-side in the overall environmental impact of online services.

\begin{enumerate}
  \item \textbf{Collect Data from ICT Sector}: We rely on the estimations provided in~\cite{aslanElectricityIntensityInternet2018} as it reviews several other studies' assumptions to provide a corrected estimate of the values and trend of the electricity intensity of data transfer: 0.06~kWh/GB in 2015 with a rate of change of one half every two years. Multiplied by the carbon cost of electricity and adjusted for MB, the corresponding carbon cost per megabyte of data transferred is:
  \begin{equation}
    \begin{aligned}
      C_{\text{transfer}} = 0.06 \times 10^{-3} \frac{1}{2^{2024-2015}} \times 445 = 52\; [\text{$\mu$gCO}_2\text{e}/\text{MB}]
    \end{aligned}
  \end{equation}
  \item \textbf{Estimate Network Operational Emissions}: Multiply the difference in data traffic (calculated in step 6 of the user-side assessment) by the cost of data transfer identified above. Given a measured difference in network data volume \( \Delta D_{f.u.} \) in megabytes between two service variants, the use-phase associated CO\(_2\)-equivalent emissions are:
  \begin{equation}
    \begin{aligned}
      \Delta U^{\text{network-side}}_{f.u.} = \Delta D_{f.u.} \times C_{\text{transfer}}
    \end{aligned}
  \end{equation}

\end{enumerate}

\subsection{Embodied and End-of-Life Assessment}

This step accounts for the environmental impact of manufacturing and disposing of the hardware used in delivering the services. We estimate the difference in embodied and EoL emissions of the two services :

\begin{enumerate}
  \item \textbf{Estimate User Device Embodied Emissions}:  
  Following the Green Metrics Tool (GMT) approach~\citeTools{greencoding}, we allocate a fraction of the user device’s total embodied emissions to a functional unit based on its execution time. To compare two services, we compute the difference in the time taken to perform the same functional unit, and use it to estimate the difference in user-side embodied emissions:
  \begin{equation}
    \begin{aligned}
    \Delta M^{user-side}_{f.u.} = M_{device}^{tot} \times \frac{\Delta t_{f.u.}}{T_{life}} \times RS
    \end{aligned}
  \end{equation}
  where:
  \begin{itemize}
    \item \( M_{device}^{tot} \): total embodied emissions of an average laptop---We use 200 kgCO\(_2\)e, as reported by L\"ovehagen et al.~\cite{lovehagen_assessing_2023}.
    \item \( \Delta t_{f.u.} \): difference in time to execute the functional unit (from step 6 of the user-side assessment).
    \item \( T_{life} \): expected device lifetime---We use 4.5 years, following manufacturer data~\cite{hp_average_lifespan}.
    \item \( RS \): resource share---Set to 1 if the device hardware is used exclusively, i.e., not shared or virtualised.
  \end{itemize}
  \item \textbf{Estimate Network Embodied Emissions}:  
  For the network infrastructure, we estimate the share of embodied emissions attributable to the functional unit based on its measured network use-phase emissions. This is done by applying a global embodied-to-use emissions ratio derived from ICT lifecycle analyses:
  \begin{equation}
    \begin{aligned}
    \Delta M^{network-side}_{f.u.} = r_{emb/use} \times \Delta U^{network-side}_{f.u.}
    \end{aligned}
  \end{equation}
  where:
  \begin{itemize}
    \item \( \Delta U^{network-side}_{f.u.} \): difference in network use-phase emissions attributable to the functional unit (from step (2) of the network-side assessment).
    \item \( r_{emb/use} \): average embodied-to-use emissions ratio for
network equipment---We use a ratio of 0.21 based on the table 8 of the
ICT-sector lifecycle analysis from Malmodin et al.~\cite{malmodin2024ict}.
  \end{itemize}
  This method assumes that embodied emissions scale proportionally with usage, which is a simplification. However, it provides a practical estimate when we consider data transfers at scale.
\end{enumerate}


\section{Experimental Setup}

\subsection{Software Tools}

In our experiments, we used several software tools to create the scenario
environment and automate user interactions while monitoring the energy
profile of the functional units. Our configurations and data are available
at~\citeTools{artifact-jason}. Whenever possible, we used open-source tools to
ensure transparency and reproducibility:
\begin{itemize}
  \item \textbf{Chromium~\citeTools{OverviewChromium}}: A popular open-source
        web browser.
        We created two profiles in Chromium: one with tracking and cookies enabled,
        and one with tracking and cookies disabled. This was used in combination
        with an ad-blocker to assess the impact of advertisement and tracking
        mechanisms on energy consumption.
  \item \textbf{Selenium~\citeTools{selenium}}: An open-source tool for
        automating web browsers, which we use to repeatedly emulate user
        interactions with the online services, such as accessing a web page,
        logging in and out, reading, sending, and deleting emails.
  \item \textbf{Pi-Hole~\citeTools{OverviewPiholePihole}}: An open-source
        DNS-based ad-blocker that blocks ads at the network level by providing
        non-routable DNS entries for ad servers. We used Pi-Hole to evaluate the
        impact of ad-blocking on the energy consumption of the online services.
  \item \textbf{Green Metric Tool (GMT)~\citeTools{greencoding} }: The GMT is
        specifically designed to monitor the power consumption of software.  By
        isolating the target program in a container, GMT creates a controlled
        environment that accurately evaluates the specific energy consumption and
        data transfer attributable to the container. The GMT collects
        energy-related data through its metric providers such as RAPL, which is a
        processor feature to gather information from system sensors. This data is
        stored in a database, which can be queried via an API. The GMT also
        includes a front-end for quick data visualisation, where it automatically
        provides insights into the emission profile. For instance, it can convert
        recorded data traffic into network emission estimates or use predefined
        machine parameters to allocate to the process under test its share of the
        machine's embodied emissions, corresponding to network and embodied
        emission assessments in our framework.
\end{itemize}

For the self-hosted email solution, we use:

\begin{itemize}
  
  \item \textbf{Roundcube~\citeTools{OverviewRoundcube}}: An open-source,
        web-based email client that offers basic email functionalities, including
        authentication, sending, replying, deleting, and filtering emails.
  \item  \textbf{Dovecot~\citeTools{OverviewDovecot}}: An open-source IMAP and
        POP3 server that acts as a mail storage server and backend for Roundcube.
  \item \textbf{Postfix~\citeTools{OverviewPostfix}}: An open-source mail
        transport agent that delivers email using SMTP. Both Postfix and Dovecot
        are part of a docker-mailserver image, which constitutes our email server
        solution in a self-hosted environment.
  \item \textbf{Mailvelope~\citeTools{OverviewMailvelope}}: A browser extension
        that adds end-to-end security through PGP encryption to webmail services.
        It integrates directly with Roundcube, enabling users to encrypt and
        decrypt messages right in the webmail interface.
  \item \textbf{TLS}: Same as with commercial providers, all  traffic in
        our self-hosted solution is encrypted using TLS to ensure secure
        communication between the client and the servers. This includes securing
        \emph{(i)} email retrieval via IMAP and POP3, \emph{(ii)} email
        transmission over SMTP, and \emph{(iii)} webmail access through HTTPS.
        
\end{itemize}

\begin{table*}[ht]
  \centering
  \caption{Pi-Hole ad-blocking effect on popular websites.}
  \vspace{-1em}
  \includegraphics[width=\textwidth]{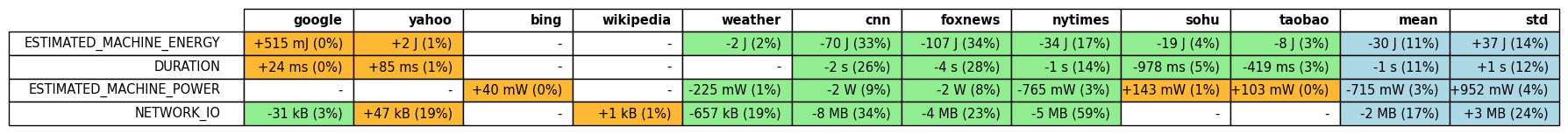}
  \label{tab:topnews_effect_of_using_adblock}
\end{table*}

\subsection{Data Collection}

Our framework involves collecting various metrics during the execution of
the functional units. The data collection process consists of the following
steps:
\begin{itemize}
  
  \item \textbf{Functional Unit Duration}: The Selenium script logs the
        start and end times of each functional unit, allowing us to measure the
        time taken to complete each task.
  \item \textbf{Energy Consumption and Network Traffic}: We used the Green
        Metric Tool (GMT)~\citeTools{greencoding} to monitor the energy consumption of
        the CPU, memory, and the entire machine. The latter is an estimation
        provided by the metric provider,
        \texttt{PsuEnergyAcXgboostMachineProvider}, installed as a submodule of the
        GMT. This module provides a model trained to estimate the AC energy
        consumption of the whole machine based on the real-time CPU utilisation by the container, and
        the machine specifications. The GMT also tracks the data traffic generated
        by the orchestrated containers.
  \item \textbf{Mean Power Consumption}: Since we deal with
        asynchronous operations---such as page loading---the duration of data
        transfers from the application servers to the browser impacts the
        energy profile. To account for this and provide a fair comparison between
        different providers, we also compute the mean power consumption from the
        energy consumption and duration metrics over the same periods.
        
\end{itemize}
Collecting data involves mapping each functional unit execution period to
its corresponding energy and traffic data using a Jupyter
notebook~\citeTools{artifact-jason}. This notebook fetches data from the GMT
database, to allow greater control than the GMT visualisation front-end. We
ensure reliability by checking the absence of error during the functional
units executions and removing outliers using the Interquartile Range (IQR)
method to filter anomalies caused by background processes. For each test we
retain at least 100 valid samples. Statistical tests (including t-tests and
normality checks) assess significant differences between test pairs. This
approach ensures that our data is accurate and reliable, providing a safe
foundation for the analysis.

\section{Experimental Results}

\subsection{Validating the Tool}\label{sec:validation}

To validate our framework, we replicated the experiments conducted by
Pearce et al.~\cite{pearce2020energy}, which compare page loading times of
websites with and without ad-blockers enabled as the basis
for energy savings estimation.
We selected the same set of popular websites and
defined a unique functional unit consisting of the user visiting the
website and waiting for the page to load entirely. We automated the process
using Selenium~\citeTools{selenium} and ran the experiments with and without
Pi-Hole ad-blocker enabled for a hundred iterations each. The results are
presented in Table~\ref{tab:topnews_effect_of_using_adblock}.

While specific results per website are presented, the mean and standard
deviation (highlighted in blue) are the values we use to validate our
tool's results in comparison to those in~\cite{pearce2020energy}. We indicate
decreases in energy, power, duration, and I/O with green and increases
with orange. The observed increase with using Pi-Hole is likely due
to the ad-blocker querying deny lists and allow lists, leading to higher
energy consumption on some websites. However, further
investigation is needed, which we leave for future work. We use the same
colour-coding scheme for tables following below.

The data in Table~\ref{tab:topnews_effect_of_using_adblock} shows that the page loading times are
significantly reduced when using Pi-Hole, with a mean reduction of 1.2
seconds across all websites. This corresponds to a 10.66\% reduction, while
the reductions reported by Pearce et al.~\cite{pearce2020energy} are 11.0\%
for AdBlock, 22.2\% for Badger Percent, and 28.5\% for uBlock Origin. Our results are consistent with those of~\cite{pearce2020energy}, validating our tool's effectiveness in measuring the impact of ad-blockers
on web page performance and supports the reliability of our tool for
assessing the environmental impact of advertisement-related content.

In their study, Pearce et al.~\cite{pearce2020energy} convert the measured time gains into corresponding energy savings using an estimate of the average laptop power consumption: 32W. Our framework goes beyond this limit and offers direct measures of the mean energy, the mean power and the generated data traffic during each functional unit. Table~\ref{tab:topnews_effect_of_using_adblock} includes those results under rows \texttt{ESTIMATED\_MACHINE\_ENERGY},
\newline \texttt{ESTIMATED\_MACHINE\_POWER} and \texttt{NETWORK\_IO}.
In contrast with the 32W reported by Pearce et al.~\cite{pearce2020energy}, we see that on average---over each website---the power consumption is 24.24W when Pi-Hole is disabled. Using Pi-Hole, further reduces the power by 0.72W on average, which corresponds to a 2.97\% reduction. Compounded with the page loading time reduction, this represents significant energy savings of 29.66J (11.36\%) per visit on average.
Additionally, the network IO are reduced by 2.48MB (16.94\%) on average. Not only this drives less network emissions, it also reflects in the machine power reduction.

\subsection{Scope and Functional Units}

We present a first application of our framework to estimate and
compare the environmental footprint of three major email providers and a self-hosted email solution:
Microsoft Outlook~\citeTools{outlook}, Gmail~\citeTools{gmail}, Proton Mail~\citeTools{proton} and the self-hosted solution~\citeTools{artifact-jason}. We aim to provide the user-side and network-side operational and embodied emission profiles for these services and assess the impact of \emph{(i)} using a given provider instead of another, \emph{(ii)} browsing with and without an ad-blocker, and \emph{(iii)} using Pretty Good Privacy (PGP) encryption. Server-side assessment is left out of the scope.

We first created different email accounts with each provider. We defined seven functional units:
\texttt{Login}, \texttt{Logout}, \texttt{No attachment},
\texttt{Attachment}, \texttt{Read}, \texttt{Reply}, and \texttt{Delete}.
The idea is to profile energy expenses, which
in turn allows estimating consumption for more complex functional units by
aggregating the measurements of basic ones. Additionally, we explicitly
defined a functional unit \texttt{Session} as the composition of
\texttt{Login} + \texttt{No attachment} + \texttt{Attachment} +
\texttt{Read} and \texttt{Reply} + \texttt{Read} and \texttt{Delete} +
\texttt{Logout}. This functional unit is used to assess the energy
consumption of a typical email session.

We then developed an automation tool using Selenium~\citeTools{selenium} to
run scenarios and monitor the energy consumption and data traffic of the functional
units. We conducted a hundred valid runs for each functional unit, with and
without the ad-blocker enabled, and with and without PGP encryption for the
self-hosted email solution. GMT recorded data on \emph{(i)} CPU, memory, and
machine energy, \emph{(ii)} network IO and the automation script tracked and recorded the \emph{(iii)} functional units' execution time.
We computed the mean CPU, memory, and
machine power consumption for each functional unit and performed a
statistical analysis to determine if there was a significant difference
between two conditions (i.e., two different providers, the same provider
with and without the ad-blocker or with and without PGP encryption).

\subsection{Assessment of Different Email Solutions}

\begin{figure}[h]
  \centering
  \includegraphics[width=\linewidth]{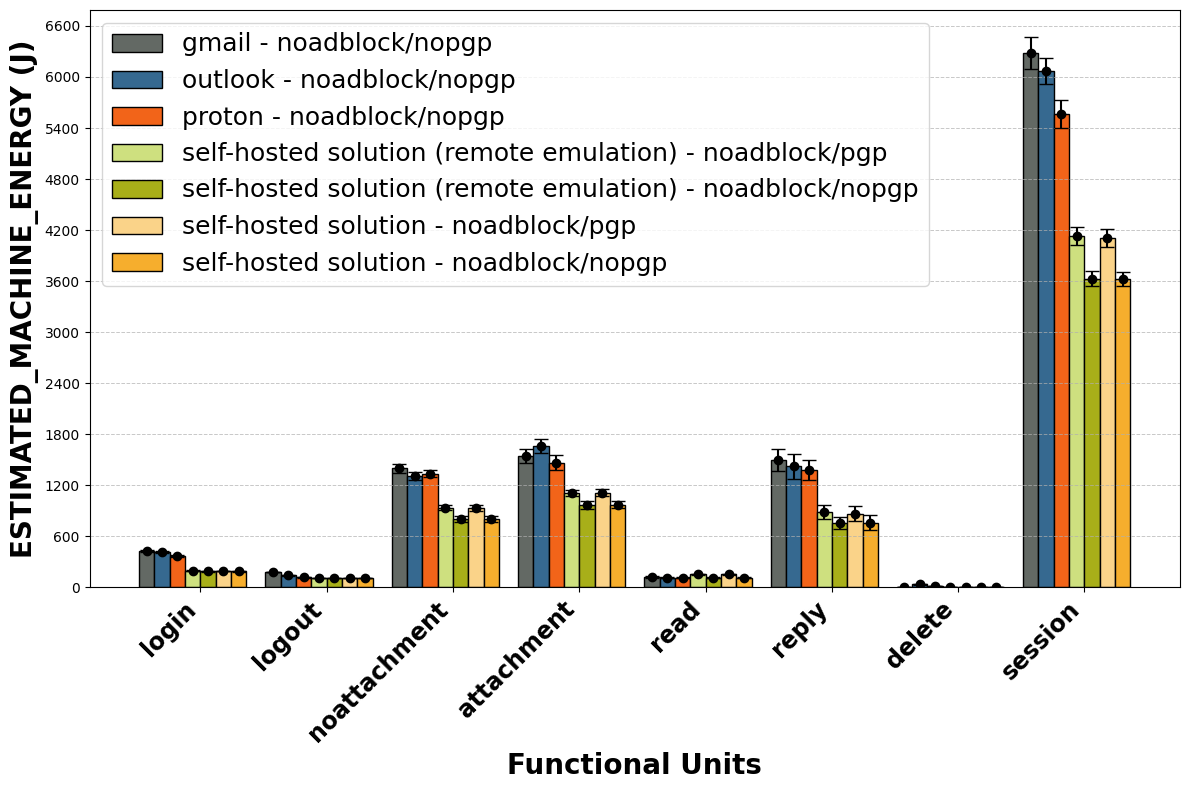}
  \caption{Estimated machine energy (J) by functional unit, without ad-blocker and self-hosted solution with PGP}
  \label{fig:machine_energy_bare_providers}
\end{figure}

We compare the energy consumption of the three email providers plus our self-hosted solution with PGP enabled. Figure~\ref{fig:machine_energy_bare_providers} provides a first insight into the energy profile for the different solutions.
It shows that the most expensive functional units are the ones associated with sending an email: \texttt{Attachment}, \texttt{NoAttachment} and \texttt{Reply}.
This is expected as these operations require more processing, user interaction and data transfer, particularly when an attachment is involved. The \texttt{Login}, \texttt{Logout}, \texttt{Read} operations are less energy-intensive while \texttt{Delete} operations appear almost negligible compared to the others.

It is also clear that the self-hosted solution is the most energy-efficient, consuming the least energy across all functional units. This result is somewhat unexpected, as we might assume the self-hosted solution to be more energy-intensive due to \emph{(i)} the resource-heavy PGP encryption operations and \emph{(ii)} the entire email system running on a single machine. Over a whole session, Gmail's webmail is the provider's front-end requiring the most energy at 6.281 kJ, followed by Outlook at 6.072 kJ, Proton Mail at 5.563 kJ, and the self-hosted solution at 4.104 kJ (4.128 kJ in remote simulation, i.e., artificial network delay).

\begin{table}[ht]
  \centering
  \caption{Mean time (ms) to ping the different provider's webmail services and our self-hosted solution}
  \vspace{-1em}
  \includegraphics[width=\linewidth]{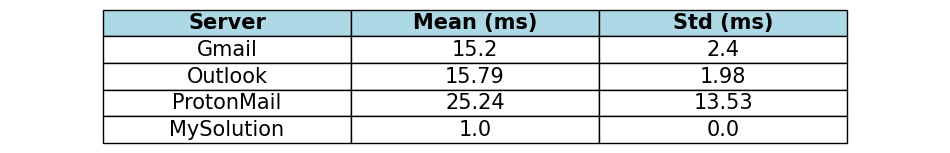}
  \label{fig:pingtest}
\end{table}

To ensure a fair comparison, we must consider an additional
factor: latency. On the client side, the time spent waiting for a web page to load contributes
to the total energy consumption, meaning providers with slower response
times might appear less efficient simply due to higher wait times. To
quantify these differences, we performed 100 ping tests for each provider,
with the results shown in Table~\ref{fig:pingtest}.
As expected, the self-hosted solution exhibits negligible latency compared to the other providers, since it operates locally. Among the main providers, Proton Mail shows the highest latency, whereas Gmail and Outlook have similar response times. This suggests that Proton Mail’s higher latency may slightly increase its measured energy consumption. If Proton Mail improved on its response time, it could potentially achieve even greater energy efficiency. Given this consideration, we can safely conclude that Proton Mail is the most energy-efficient option among the three main providers.

To fairly compare the self-hosted solution with the other providers, we repeated the same series of tests, but this time added a simulated 50 ms network delay. This artificially emulates a realistic remote access scenario, such as working from home. The results, also shown in Figure~\ref{fig:machine_energy_bare_providers}, show that the energy consumption of the self-hosted solution remains nearly unchanged. It continues to be the most energy-efficient option overall.

It’s worth noting that running a self-hosted solution on every user’s
personal device would not be practical, as it would require each user machine to stay powered on at all times---leading to much higher emissions. A more realistic and sustainable option would be for small organisations, such as universities, to host their own email services for members. This setup not only improves data sovereignty but also reduces emissions. In this particular case, because the whole solution is running on the same machine, both server- and client-side emissions are included in the operational user-side assessment, which further highlights the advantage of the self-hosted solution, already the lowest in energy use.

\renewcommand{\tablename}{Tables}
\begin{table*}[ht]
  \centering
  \caption{Comparison of email solutions and their emission savings}

  \begin{subfigure}{\textwidth}
    \centering%
    \caption{Effect of using Proton Mail instead of Outlook (no ad-blocker)}
    \vspace{-1mm}
    \includegraphics[width=\textwidth]{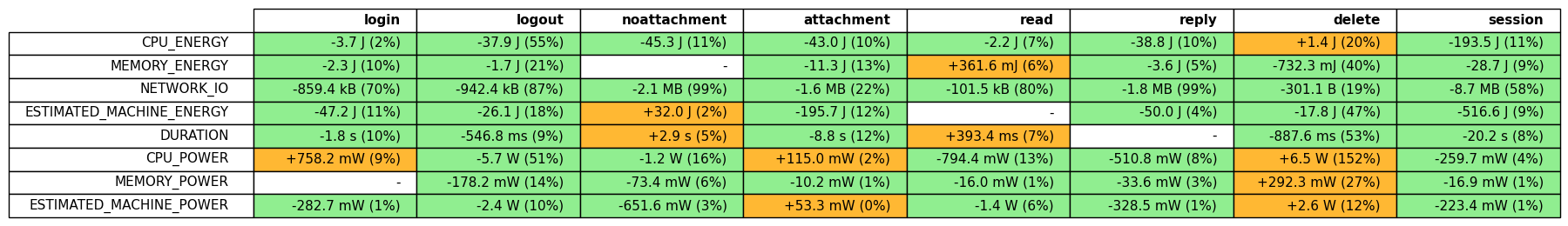}
    \label{tab:proton_vs_outlook}
  \end{subfigure}

  \begin{subfigure}{\textwidth}
    \centering%
    \caption{Emission savings (CO2-eq.) using Proton Mail instead of Outlook (no ad-blocker)}
    \vspace{-1mm}
    \includegraphics[width=\textwidth]{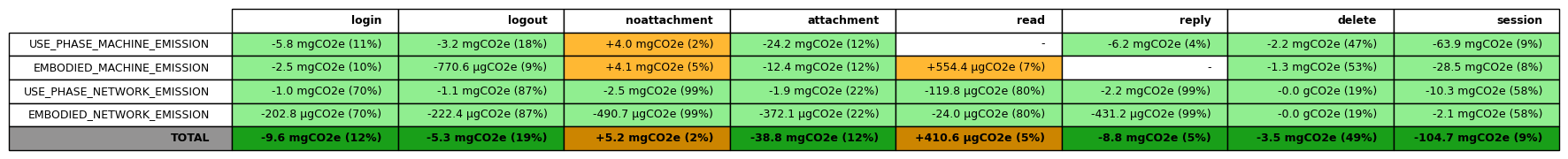}
    \label{tab:proton_vs_outlook_co2}
  \end{subfigure}

  \begin{subfigure}{\textwidth}
    \centering%
    \caption{Effect of using the self-hosted solution remotely with PGP instead of Gmail (no ad-blocker)}
    \vspace{-1mm}
    \includegraphics[width=\textwidth]{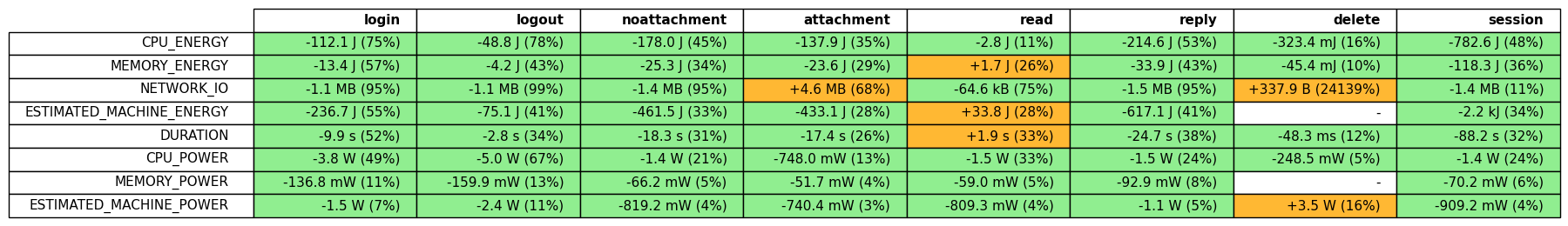}
    \label{tab:mysolution_vs_gmail}
  \end{subfigure}

  \begin{subfigure}{\textwidth}
    \centering%
    \caption{Emission savings (CO2-eq.) using a self-hosted solution remotely with PGP instead of Gmail (no ad-blocker)}
    \vspace{-1mm}
    \includegraphics[width=\textwidth]{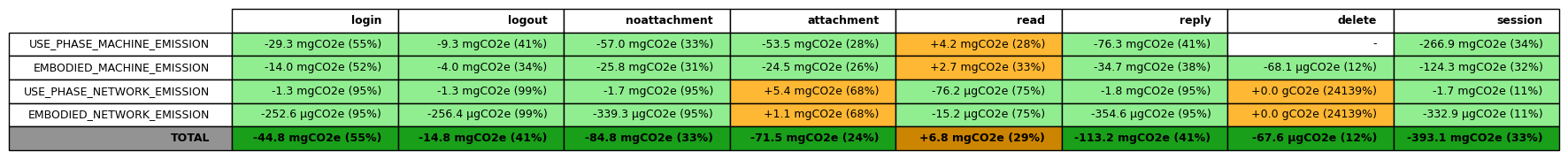}
    \label{tab:mysolution_vs_gmail_co2}
  \end{subfigure}

  \begin{subfigure}{\textwidth}
    \centering%
    \caption{Emission savings (CO\(_2\)-eq.) migrating all Gmail active users (one session per month) to remote self-hosted solutions without PGP}
    \vspace{-1mm}
    \includegraphics[width=\textwidth]{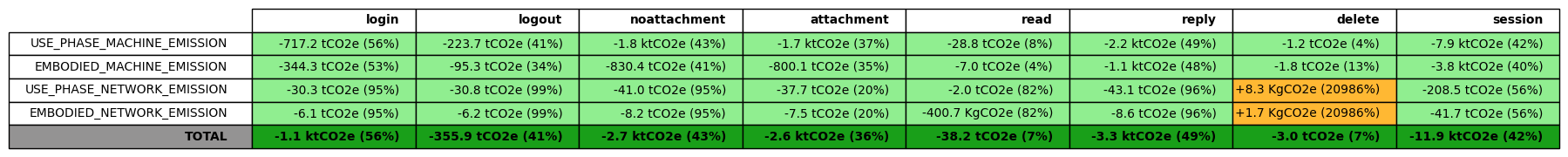}
    \label{tab:mysolution_vs_gmail_all_users}
  \end{subfigure}
\end{table*}
\renewcommand{\tablename}{Table}

\begin{figure}[H]
  \centering
  \includegraphics[width=\linewidth]{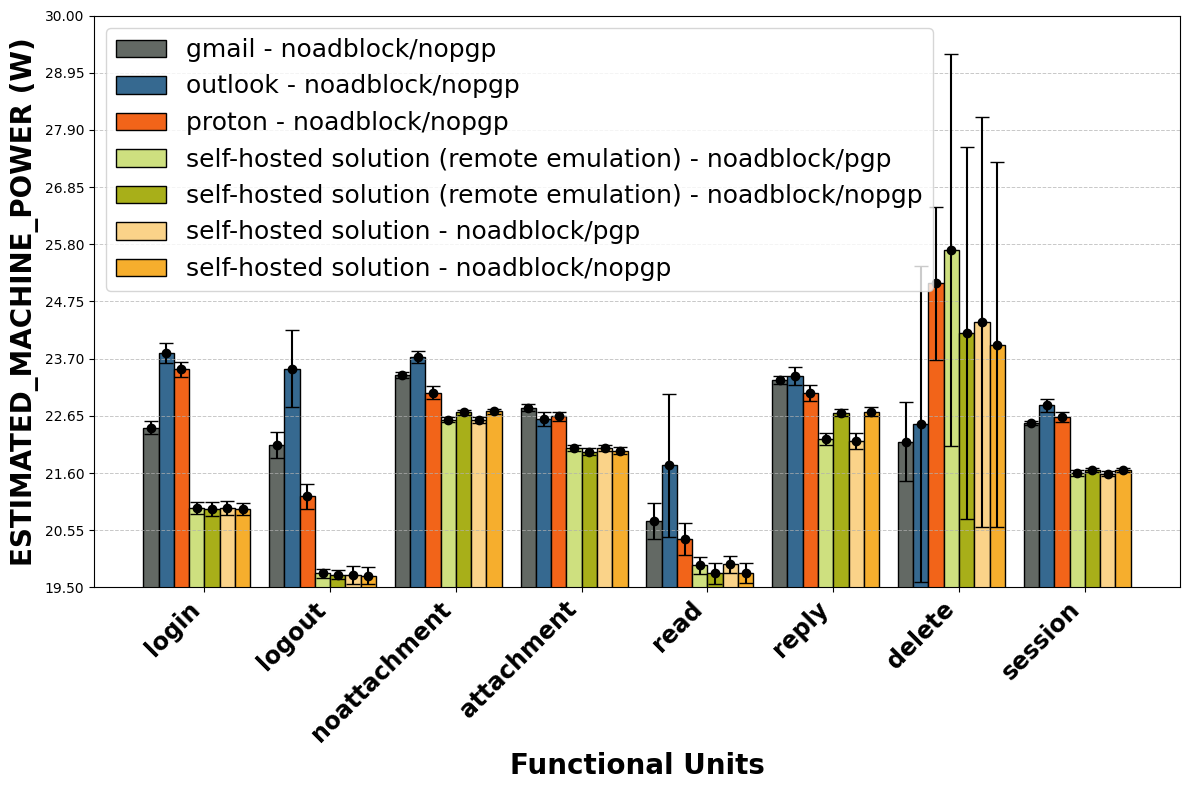}
  \caption{Estimated machine power (W) by functional unit, without ad-blocker and self-hosted solution with PGP}
  \label{fig:machine_power_bare_providers}
\end{figure}

Next, we examine the mean power profile provided in Figure~\ref{fig:machine_power_bare_providers}. The self-hosted solution is more power-efficient than the main providers, with the lowest mean power consumption for all functional units including the entire session. An exception is the \texttt{delete} operation, but it can be ignored, as this functional unit uses a negligible amount of energy overall. In addition, the self-hosted solution is also the fastest among all providers. As shown in Table~\ref{tab:mysolution_vs_gmail}, a session runs 32\% faster on the self-hosted solution compared to Gmail. This highlights a two-fold advantage of the self-hosted setup: it runs faster and is more power-efficient.

\begin{figure}[H]
  \centering
  \includegraphics[width=\linewidth]{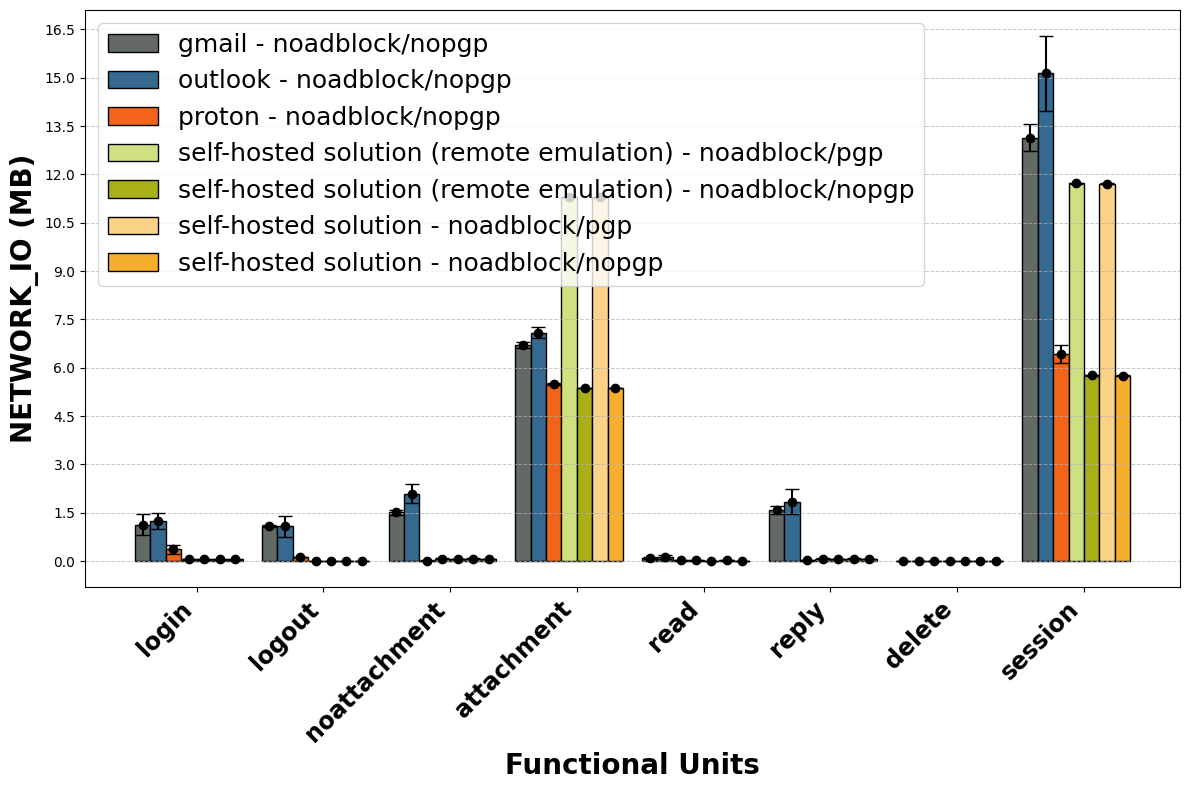}
  \vspace{-1em}
  \caption{Network IO (MB) by functional unit, without ad-blocker and self-hosted solution with PGP}
  \label{fig:network_io_bare_providers}
\end{figure}

Another interesting result is provided in Figure~\ref{fig:network_io_bare_providers}, which shows the network traffic generated by the different providers. Over the entire session, Gmail and Outlook generated more than twice the data traffic of Proton Mail and the self-hosted solution without PGP. This result is important as it uncovers a fundamental difference between privacy-focused services, such as Proton Mail and the self-hosted solution, and advertising- and tracking-supported providers like Gmail and Outlook.
The latter inherently require more data transfer to support tracking, personalisation, and ad delivery, which directly contributes to higher network and device energy consumption.
While using PGP, the self-hosted solution generates more than 11MB to send an email with 5MB attachment---part of the encrypted content---which is more than Proton but still less than Outlook and Gmail. This is because PGP encryption produces binary output, which must be encoded in Base64 to be transmitted via SMTP. Base64 encoding increases the size of the original data by approximately 33\%, contributing significantly to the overall increase in traffic. This result highlights the trade-off between security and energy consumption. It is important to note though, that the entire session's traffic is still lower than that of Gmail and Outlook, while Proton Mail barely exceeds its 5MB of attachment. As we already argued, the network traffic, at scale, is a key driver of network capacity which is in turn a driver of emissions.
Although the linearity of the relation is questionable, Tables~\ref{tab:proton_vs_outlook_co2}  and~\ref{tab:mysolution_vs_gmail_co2} present corresponding CO\(_2\)e network emissions savings assuming an effective difference in network capacity requirements for using email services.

These findings suggest that the choice of email provider can significantly impact the energy consumption and emissions of a user device. To illustrate this, Tables~\ref{tab:proton_vs_outlook} and~\ref{tab:mysolution_vs_gmail} show the reduction in energy, execution time and data consumption when using Proton Mail instead of Outlook and the self-hosted solution with PGP instead of Gmail. These comparisons highlight the potential energy savings achievable by switching to a more efficient provider.
Similarly, Tables~\ref{tab:proton_vs_outlook_co2} and~\ref{tab:mysolution_vs_gmail_co2} present the corresponding reductions in CO\(_2\)-equivalent emissions, providing further insight into the environmental benefits of these choices. Specifically, using Proton Mail instead of Outlook can leverage up to 0.1gCO\(_2\)e per session (9\% of Outlook's overall user and network sides emissions), while the self-hosted solution instead of Gmail can leverage up to 0.39gCO\(_2\)e per session (33\% of Gmail's overall user and network sides emissions).

We insist on the fact that while the self-hosted solution shows significant energy savings in our tests, this result should be interpreted with caution. If each user were to run their own mail server, the overall energy consumption would likely increase dramatically, as these servers need to be constantly running. Instead, as previously discussed, our findings suggest that shared self-hosting by small organisations can offer an ideal compromise that also promotes data sovereignty. Furthermore, there is room for energy consumption improvements at major email providers, even when considering the additional layer of PGP encryption. Overall, these results highlight the potential for both large-scale providers and smaller institutions to optimise email infrastructure for better energy efficiency and reduced environmental impact.

To further illustrate the potential scale of these savings, consider Gmail’s estimated 2 billion active users~\cite{kemp_digital_2025}, each performing just one session per month. If all these users switched to remote self-hosted solutions without PGP encryption, the resulting annual savings could reach up to 11.9 ktCO\(_2\)e, a 42\% reduction in Gmail’s user and network sides emissions, as shown in Table~\ref{tab:mysolution_vs_gmail_all_users}. This is equivalent to the emissions from approximately 9,000 round-trip flights between Paris and New York. Similarly, as detailed in our public repository~\citeTools{artifact-jason}, migrating all Outlook users to Proton Mail could yield annual savings of up to 2.2 ktCO\(_2\)e.

\subsection{Assessment of Ad-Blocker}

We evaluate the impact of ad-blocking on the energy consumption of email services. Among the tested providers, Outlook is the only one where ad-blocking yields an appreciable effect, as it is the only service displaying ads in our setup. For the other providers, we observed a slight increase in session duration and energy consumption when using Pi-Hole, which we attribute to the additional time required for scanning its list of non-routable DNS entries.
Table~\ref{tab:adblock_outlook}
shows the energy, execution time and data savings achieved with Pi-Hole, while
Table~\ref{tab:co2_adblock_outlook} presents the corresponding
CO\(_2\)-equivalent emission reductions. The interpretation is straightforward: by blocking ads, Pi-Hole reduces both the amount of processed data and the time spent loading content. On average, it decreases session duration by 3.69s (1.39\%), machine power consumption by 0.13W (0.55\%), and total energy consumption by 117.35J (1.93\%), leading to an emission reduction of 0.02gCO\(_2\)e per session (2\% of Outlook's overall user and network sides emissions).

As expected, the network traffic generated is also reduced by 1.27MB (8.4\%). Given Outlook's nearly 400 million active users, this could translate into emission savings of 448 tonnes of CO\(_2\)e annually, assuming a hypothetical but reasonable average usage frequency of one session per week (Table~\ref{tab:co2_adblock_outlook_all_users}). These findings align with those of Pearce et al.~\cite{pearce2020energy}, who reported that ad-blockers can reduce page loading times by up to 28\%, potentially saving 13.5 billion kWh per year globally. Our results suggest that reasonable energy savings could be achieved on email services displaying ads.

\subsection{Assessment of PGP Encryption}

We also assess the impact of PGP encryption on the energy consumption of our self-hosted solution. PGP encryption is a widely used method for securing email communications, but it can also increase energy consumption due to the additional processing required for encrypting and decrypting messages. To evaluate this impact, we compared the energy consumption of the self-hosted solution with and without PGP enabled. As expected, PGP encryption increases both energy consumption (+14\%)
and data traffic (+103\%) for the self-hosted solution~\citeTools{artifact-jason}. This is due to the additional processing required for encrypting and decrypting emails. The effect concerns only the functional units \texttt{Attachment}, \texttt{No attachment}, \texttt{Read}, and \texttt{Reply}, which involve encryption or decryption. Overall, session energy emissions rises by 0.1gCO\(_2\)e per session, corresponding to 15\% of the self-hosted solution overall user and network sides emissions, as detailed in Table~\ref{tab:pgp_effect}.
This result highlights the natural trade-off between security and energy efficiency.
\renewcommand{\tablename}{Tables}
\begin{table*}

  \centering
  \caption{Overview of Pi-Hole and PGP effects on Outlook and self-hosted solutions.}
  
  \begin{subfigure}{\textwidth}
    \centering%
    \caption{Pi-Hole ad-blocking effect on Outlook}
    \vspace{-1mm}
    \includegraphics[width=\textwidth]{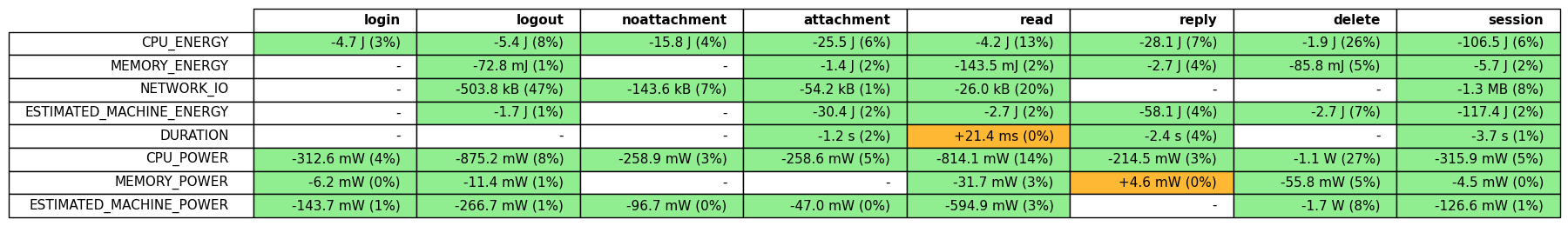}
    \label{tab:adblock_outlook}
  \end{subfigure}
  \vspace{1em}
  \begin{subfigure}{\textwidth}
    \centering%
    \caption{Emission savings (CO\(_2\)-eq.) activating Pi-Hole on Outlook}
    \vspace{-1mm}
    \includegraphics[width=\textwidth]{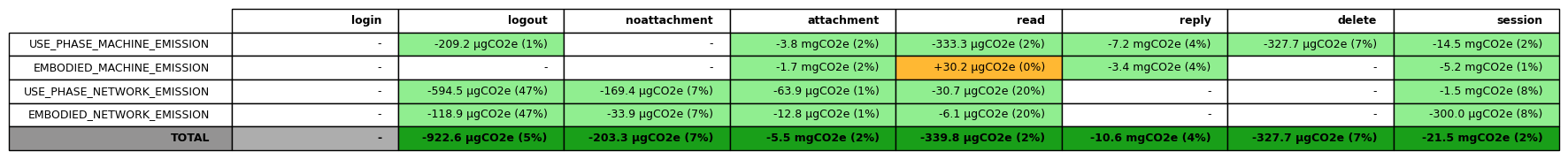}
    \label{tab:co2_adblock_outlook}
  \end{subfigure}
  \vspace{1em}
  \begin{subfigure}{\textwidth}
    \centering%
    \caption{Emission savings (CO\(_2\)e) activating Pi-Hole for all Outlook active users, assuming one session per week (hypothetical)}
    \vspace{-1mm}
    \includegraphics[width=\textwidth]{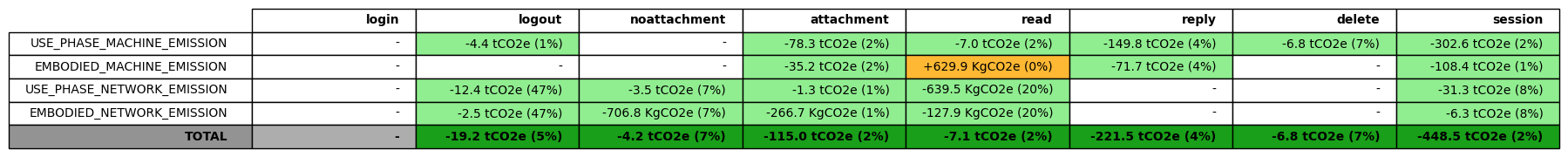}
    \label{tab:co2_adblock_outlook_all_users}
  \end{subfigure}
  \vspace{1em}
  \begin{subfigure}{\textwidth}
    \centering%
    \caption{PGP encryption effect on the self-hosted solution}
    \vspace{-1mm}
    \includegraphics[width=\textwidth]{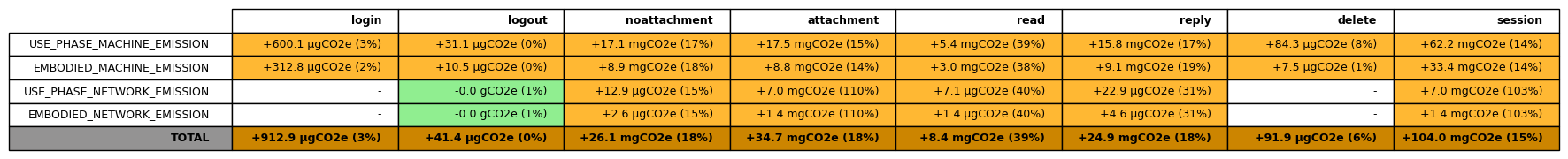}
    \label{tab:pgp_effect}
  \end{subfigure}
\end{table*}
\renewcommand{\tablename}{Table}

\FloatBarrier
\section{Limitations and Future Directions}

The primary limitation of this work is its scope. It provides a way to estimate user and network sides emissions of online services but misses including server-side emissions.
Additionally, While the strength of this work is in the use of a bottom-up approach to estimate user-side emissions, the network emissions are derived from secondary data on broader ICT sector, using measured traffic volume as an allocation criterion. In effect, this assumes elastic network emissions with respect to traffic volume, an assumption which has its limitation.

\subsection{Server-Side Assessment}

Estimating server-side emissions is challenging due to the lack of transparency regarding environmental impacts, and data processing and storage from companies handling user information. These emissions are expected to be substantial for ad-supported services. In particular, the training and maintaining of advertising models is assumed to have high impacts that are hard to assess: training AI models is not only known to be a computationally intensive task but also often requires extra hardware accumulating thereby more embodied emissions.

Previous studies~\cite{freitag_real_2021, malmodin2024ict} assessing the ICT sector environmental impact often rely on global market and industry statistics. These high-level estimates come with significant uncertainty and lack service-specific granularity. We initially considered allocating these server-side emissions estimates proportionally to the volume of data transferred. However, this method fails to account for the nature of the underlying data processing: it assigns identical emissions per unit of data to all services, regardless of whether the data is merely stored or used to train complex machine learning models. While monitoring network traffic provides an estimate of the data volume collected by these platforms, we still lack the energy intensity of the computation happening behind the scenes, simply because we do not know what type of computation that is. As a result, we chose not to include server-side estimates based on this allocation strategy, as they would be inaccurate and potentially misleading. Instead, we focus on providing reliable user-side and network-side emissions estimates.

We acknowledge that this is a significant limitation of our work, as it prevents us from providing a complete picture of the environmental footprint of online services. An extension of the proposed framework to include server-side emissions could, e.g., listing various data storage and processing technologies with estimates of their carbon cost per gigabyte (cf.~\cite{rao_understanding_2024}, for example). However, such insights would remain indicative at best and could not serve as a basis for direct comparison. Without regulations requiring stakeholders to disclose much more detailed information about their activities, it will remain challenging to produce comprehensive and accurate estimates for specific service operators.

\subsection{Network-Side Assessment}

In terms of network consumption, we believe our estimates are closer to reality. However, as Freitag et al. note in their comparative analysis of ICT climate impact estimates~\cite{freitag_real_2021}, the linear data volume-based allocation of emissions remains contested. We recognise that network consumption is not elastic: it is designed for peak capacity and consumes a relatively constant amount of energy regardless of traffic. Still, since increasing data demand eventually drives infrastructure expansion and higher emissions, we maintain that assigning responsibility based on the volume of data exchanged remains a reasonable approximation~\cite{aslanElectricityIntensityInternet2018}.

For embodied emissions, we apply a global embodied-to-use ratio derived from ICT lifecycle analyses. Although this simplifies estimation, it inherits the same limitation: it distributes a mostly fixed impact according to variable usage. This may give a false impression---for example, that reducing individual traffic could lower emissions, when in fact emissions remain constant unless demand is reduced at scale. At best, lower usage helps avoid future increases. Despite these constraints, proportional allocation remains a practical heuristic for comparison in the absence of more detailed models.

\subsection{User-Side Assessment}

On the user side, for which we have the most reliable estimates, a few limitations need to be mentioned. Webmail services typically display few ads but rely on data collection and reselling as their business model. Ad-blockers are less effective against such mechanisms and most processing happens server-side. This could explain the lack of impact measured on Gmail in our results. We are also critically aware of projections claiming that between 50\,\% and 90\,\% of all email traffic is spam~\cite{statistaspam2024,fu2014detecting}, the management of which might have a substantial impact on sustainability metrics. We leave assessment and evaluation of these aspects for future work.


\section{Conclusions}

Our work offers insights into the environmental impact of online services. We proposed an approach and framework for the impact assessment of software systems and applied it to evaluate the user-side and network-side emissions of webmail services. Our approach not only facilitates comparison between similar services based on energy consumption, but also highlights the connection between data collection, processing, and their environmental costs. The results of our study support our hypothesis that \emph{online services that emphasise privacy and security are typically \enquote{greener.}}

Quantitatively, our experiments show that using Proton Mail instead of Outlook can reduce emissions by up to 0.1 gCO\(_2\)e (9\%) per session. More significantly, switching from Gmail to our self-hosted solution with PGP encryption results in savings of up to 0.39 gCO\(_2\)e (33\%) per session. We also evaluated the impact of end-to-end encryption: enabling PGP increases energy consumption by 13.46\%, data traffic by 103.34\% and emissions by 15\% compared to the same self-hosted setup without PGP. This overhead is mostly due to the computational cost of encryption and the increased payload size from Base64 encoding. Without PGP and scaled to the total number of active Gmail users, the emissions savings could reach up to 11.9 ktCO\(_2\)e annually (a 42\% reduction in Gmail’s user-side and network-side emissions), which is equivalent to the emissions from approximately 9,000 round-trip flights between Paris and New York.  Nevertheless, even with PGP enabled, the self-hosted setup remains the most energy-efficient of all tested configurations. 
Ad-blocking, tested with Pi-Hole, reduces Outlook's user-side emissions by 2\%, or approximately 0.02 gCO\(2\)e per session. Given Outlook’s nearly 400 million active users, this translates to potential savings of up to 448 tCO\(2\)e annually under reasonable usage assumptions. Another example involving the migration of the Outlook users to Proton Mail could yield a 2.2 ktCO\(_2\)e annual saving. These results show that seemingly small optimisations at the session level can lead to substantial cumulative benefits at scale.

While our work shows that the self-hosted solution is the most efficient–regarding user and network sides emissions–, we are not by any means advocating for every user to host their own mail server, as this will result not only in security and privacy issues but also environmental impacts due to less optimal server usage and additional hardware needs. However, we do agree with Fiebig et al. that
large institutions and public services such as universities or hospitals should host their own service. This does not only have the potential to improve on privacy, security and sustainability, but it is also a step towards digital sovereignty.  Digital sovereignty is often discussed in terms of owning data, however, we do share the opinion of Fiebig et al. that it should also include the perspective of the ability to operate, repair, and rebuild digital infrastructure instead of it being monopolised by Silicon-Valley Corporations~\cite{fiebig202213}, including popular email and cloud providers assessed in our work. As beautifully noted by Fiebig et al. in \enquote{Escaping Academic Cloudification to Preserve Academic Freedom} discussing public infrastructures in public services \enquote{when truly sustainable, the question of privacy and academic freedom will solve themselves}~\cite{fiebig2022position}.

Future work should extend this framework to include server-side emissions for a more complete picture. Nonetheless, our findings already demonstrate the empirical value of bottom-up measurements for sustainability benchmarking and confirm that privacy-driven services can also be the greener choice. However, we strongly believe that without regulations requiring stakeholders to disclose much more information about their infrastructures and processing activities, producing comprehensive and accurate environmental estimates will remain inherently limited.

\begin{acks}
  We gratefully acknowledge the Brussels-Capital Region -- Innoviris for
financial support under grant numbers 2024-RPF-2 MImPG and 2024-RPF-4 SDM, and the
CyberExcellence programme of the Walloon Region, Belgium (grant
2110186).
\hyphenation{Innoviris}

\end{acks}

\clearpage
\balance

\bibliographystyleTools{ACM-Reference-Format}
\bibliographyTools{tools}

\bibliographystyle{ACM-Reference-Format}
\bibliography{sustainable-ict}


\begin{thebibliography}{13}


\ifx \showCODEN    \undefined \def \showCODEN     #1{\unskip}     \fi
\ifx \showDOI      \undefined \def \showDOI       #1{#1}\fi
\ifx \showISBNx    \undefined \def \showISBNx     #1{\unskip}     \fi
\ifx \showISBNxiii \undefined \def \showISBNxiii  #1{\unskip}     \fi
\ifx \showISSN     \undefined \def \showISSN      #1{\unskip}     \fi
\ifx \showLCCN     \undefined \def \showLCCN      #1{\unskip}     \fi
\ifx \shownote     \undefined \def \shownote      #1{#1}          \fi
\ifx \showarticletitle \undefined \def \showarticletitle #1{#1}   \fi
\ifx \showURL      \undefined \def \showURL       {\relax}        \fi
\providecommand\bibfield[2]{#2}
\providecommand\bibinfo[2]{#2}
\providecommand\natexlab[1]{#1}
\providecommand\showeprint[2][]{arXiv:#2}

\bibitem[\protect\citeauthoryear{??}{Ove}{2024a}]%
        {OverviewChromium}
 \bibinfo{year}{2024}\natexlab{a}.
\newblock \bibinfo{title}{The Chromium Projects}.
\newblock
\newblock
\urldef\tempurl%
\url{https://www.chromium.org/}
\showURL{%
\tempurl}
\newblock
\shownote{Accessed: 2024-11-15.}


\bibitem[\protect\citeauthoryear{??}{Ove}{2024b}]%
        {OverviewDovecot}
 \bibinfo{year}{2024}\natexlab{b}.
\newblock \bibinfo{title}{DOVECOT The Secure IMAP server}.
\newblock
\newblock
\urldef\tempurl%
\url{https://www.dovecot.org/}
\showURL{%
\tempurl}
\newblock
\shownote{Accessed: 2024-10-02.}


\bibitem[\protect\citeauthoryear{??}{gma}{2024}]%
        {gmail}
 \bibinfo{year}{2024}\natexlab{}.
\newblock \bibinfo{title}{Gmail}.
\newblock
\newblock
\urldef\tempurl%
\url{https://www.google.com/gmail/about/}
\showURL{%
\tempurl}
\newblock
\shownote{Accessed: 2024-10-02.}


\bibitem[\protect\citeauthoryear{??}{gre}{2024}]%
        {greencoding}
 \bibinfo{year}{2024}\natexlab{}.
\newblock \bibinfo{title}{Green Metrics Tool}.
\newblock
\newblock
\urldef\tempurl%
\url{https://docs.green-coding.io/docs/prologue/philosophy-methodology/}
\showURL{%
\tempurl}
\newblock
\shownote{Accessed: 2024-10-02.}


\bibitem[\protect\citeauthoryear{??}{Ove}{2024c}]%
        {OverviewMailvelope}
 \bibinfo{year}{2024}\natexlab{c}.
\newblock \bibinfo{title}{Mailvelope}.
\newblock
\newblock
\urldef\tempurl%
\url{https://mailvelope.com/en}
\showURL{%
\tempurl}
\newblock
\shownote{Accessed: 2024-10-02.}


\bibitem[\protect\citeauthoryear{??}{out}{2024}]%
        {outlook}
 \bibinfo{year}{2024}\natexlab{}.
\newblock \bibinfo{title}{Microsoft Outlook}.
\newblock
\newblock
\urldef\tempurl%
\url{https://outlook.live.com/mail/}
\showURL{%
\tempurl}
\newblock
\shownote{Accessed: 2024-10-02.}


\bibitem[\protect\citeauthoryear{??}{Ove}{2024d}]%
        {OverviewPiholePihole}
 \bibinfo{year}{2024}\natexlab{d}.
\newblock \bibinfo{title}{Overview of {Pi}-hole - {Pi}-hole documentation}.
\newblock
\newblock
\urldef\tempurl%
\url{https://docs.pi-hole.net/}
\showURL{%
\tempurl}
\newblock
\shownote{Accessed: 2024-11-15.}


\bibitem[\protect\citeauthoryear{??}{Ove}{2024e}]%
        {OverviewPostfix}
 \bibinfo{year}{2024}\natexlab{e}.
\newblock \bibinfo{title}{The Postfix}.
\newblock
\newblock
\urldef\tempurl%
\url{https://www.postfix.org/}
\showURL{%
\tempurl}
\newblock
\shownote{Accessed: 2024-10-02.}


\bibitem[\protect\citeauthoryear{??}{pro}{2024}]%
        {proton}
 \bibinfo{year}{2024}\natexlab{}.
\newblock \bibinfo{title}{Proton Mail}.
\newblock
\newblock
\urldef\tempurl%
\url{https://proton.me/mail}
\showURL{%
\tempurl}
\newblock
\shownote{Accessed: 2024-10-02.}


\bibitem[\protect\citeauthoryear{??}{Ove}{2024f}]%
        {OverviewRoundcube}
 \bibinfo{year}{2024}\natexlab{f}.
\newblock \bibinfo{title}{Roundcube webmail}.
\newblock
\newblock
\urldef\tempurl%
\url{https://roundcube.net/}
\showURL{%
\tempurl}


\bibitem[\protect\citeauthoryear{??}{sel}{2024}]%
        {selenium}
 \bibinfo{year}{2024}\natexlab{}.
\newblock \bibinfo{title}{Selenium}.
\newblock
\newblock
\urldef\tempurl%
\url{https://www.selenium.dev}
\showURL{%
\tempurl}
\newblock
\shownote{Accessed: 2024-10-02.}


\bibitem[\protect\citeauthoryear{??}{ele}{2025}]%
        {electricitymap}
 \bibinfo{year}{2025}\natexlab{}.
\newblock \bibinfo{title}{Live 24/7 CO2 emissions of electricity consumption}.
\newblock
\newblock
\urldef\tempurl%
\url{http://electricitymap.tmrow.co}
\showURL{%
\tempurl}
\newblock
\shownote{Accessed: 2025-03-08.}


\bibitem[\protect\citeauthoryear{Kayembe}{Kayembe}{2024}]%
        {artifact-jason}
\bibfield{author}{\bibinfo{person}{Jason Kayembe}.} \bibinfo{year}{2024}\natexlab{}.
\newblock \bibinfo{title}{Artifact: Exploring Privacy and Security as Drivers for Environmental Sustainability in Cloud-Based Office Solutions}.
\newblock
\newblock
\urldef\tempurl%
\url{https://doi.org/10.5281/zenodo.15729970}
\showDOI{\tempurl}


\end{thebibliography}



\begin{thebibliography}{18}


\ifx \showCODEN    \undefined \def \showCODEN     #1{\unskip}     \fi
\ifx \showDOI      \undefined \def \showDOI       #1{#1}\fi
\ifx \showISBNx    \undefined \def \showISBNx     #1{\unskip}     \fi
\ifx \showISBNxiii \undefined \def \showISBNxiii  #1{\unskip}     \fi
\ifx \showISSN     \undefined \def \showISSN      #1{\unskip}     \fi
\ifx \showLCCN     \undefined \def \showLCCN      #1{\unskip}     \fi
\ifx \shownote     \undefined \def \shownote      #1{#1}          \fi
\ifx \showarticletitle \undefined \def \showarticletitle #1{#1}   \fi
\ifx \showURL      \undefined \def \showURL       {\relax}        \fi
\providecommand\bibfield[2]{#2}
\providecommand\bibinfo[2]{#2}
\providecommand\natexlab[1]{#1}
\providecommand\showeprint[2][]{arXiv:#2}

\bibitem[\protect\citeauthoryear{??}{hp_}{2024}]%
        {hp_average_lifespan}
 \bibinfo{year}{2024}\natexlab{}.
\newblock \bibinfo{title}{Average {Computer} {Lifespan}: {How} {Long} {Do} {PCs} {Last}? {\textbar} {HP}® {Tech} {Takes}}.
\newblock
\newblock
\urldef\tempurl%
\url{https://www.hp.com/us-en/shop/tech-takes/average-computer-lifespan}
\showURL{%
\tempurl}


\bibitem[\protect\citeauthoryear{??}{Eur}{2024}]%
        {EuropeanCommissionsUse2024}
 \bibinfo{year}{2024}\natexlab{}.
\newblock \bibinfo{title}{European {Commission}’s use of {Microsoft} 365 infringes data protection law for {EU} institutions and bodies {\textbar} {European} {Data} {Protection} {Supervisor}}.
\newblock
\newblock
\urldef\tempurl%
\url{https://www.edps.europa.eu/press-publications/press-news/press-releases/2024/european-commissions-use-microsoft-365-infringes-data-protection-law-eu-institutions-and-bodies}
\showURL{%
\tempurl}


\bibitem[\protect\citeauthoryear{Aslan, Mayers, Koomey, and France}{Aslan et~al\mbox{.}}{2018}]%
        {aslanElectricityIntensityInternet2018}
\bibfield{author}{\bibinfo{person}{Joshua Aslan}, \bibinfo{person}{Kieren Mayers}, \bibinfo{person}{Jonathan~G. Koomey}, {and} \bibinfo{person}{Chris France}.} \bibinfo{year}{2018}\natexlab{}.
\newblock \showarticletitle{Electricity {Intensity} of {Internet} {Data} {Transmission}: {Untangling} the {Estimates}}.
\newblock \bibinfo{journal}{\emph{Journal of Industrial Ecology}} \bibinfo{volume}{22}, \bibinfo{number}{4} (\bibinfo{year}{2018}), \bibinfo{pages}{785--798}.
\newblock
\showISSN{1530-9290}
\urldef\tempurl%
\url{https://doi.org/10.1111/jiec.12630}
\showDOI{\tempurl}


\bibitem[\protect\citeauthoryear{B.}{B.}{2018}]%
        {Wolford_gdpr_2018}
\bibfield{author}{\bibinfo{person}{Wolford B.}} \bibinfo{year}{2018}\natexlab{}.
\newblock \bibinfo{title}{What is {GDPR}, the {EU}’s new data protection law?}
\newblock
\newblock
\urldef\tempurl%
\url{https://gdpr.eu/what-is-gdpr/}
\showURL{%
\tempurl}


\bibitem[\protect\citeauthoryear{Fiebig and Aschenbrenner}{Fiebig and Aschenbrenner}{2022}]%
        {fiebig202213}
\bibfield{author}{\bibinfo{person}{Tobias Fiebig} {and} \bibinfo{person}{Doris Aschenbrenner}.} \bibinfo{year}{2022}\natexlab{}.
\newblock \showarticletitle{13 propositions on an internet for a" burning world"}. In \bibinfo{booktitle}{\emph{Proceedings of the ACM SIGCOMM Joint Workshops on Technologies, Applications, and Uses of a Responsible Internet and Building Greener Internet}}. \bibinfo{pages}{1--5}.
\newblock
\urldef\tempurl%
\url{https://doi.org/10.1145/3538395.3545312}
\showURL{%
\tempurl}


\bibitem[\protect\citeauthoryear{Fiebig, G{\"u}rses, Gañán, Kotkamp, Kuipers, Lindorfer, Prisse, and Sari}{Fiebig et~al\mbox{.}}{2023}]%
        {fiebig_heads_2023}
\bibfield{author}{\bibinfo{person}{Tobias Fiebig}, \bibinfo{person}{Seda G{\"u}rses}, \bibinfo{person}{Carlos~H. Gañán}, \bibinfo{person}{Erna Kotkamp}, \bibinfo{person}{Fernando Kuipers}, \bibinfo{person}{Martina Lindorfer}, \bibinfo{person}{Menghua Prisse}, {and} \bibinfo{person}{Taritha Sari}.} \bibinfo{year}{2023}\natexlab{}.
\newblock \showarticletitle{Heads in the {Clouds}? {Measuring} {Universities}’ {Migration} to {Public} {Clouds}: {Implications} for {Privacy} \& {Academic} {Freedom}}.
\newblock \bibinfo{journal}{\emph{Proceedings on Privacy Enhancing Technologies}} (\bibinfo{year}{2023}).
\newblock
\showISSN{2299-0984}
\urldef\tempurl%
\url{https://doi.org/10.56553/popets-2023-0044}
\showDOI{\tempurl}


\bibitem[\protect\citeauthoryear{Fiebig, Lindorfer, and G{\"u}rses}{Fiebig et~al\mbox{.}}{2022}]%
        {fiebig2022position}
\bibfield{author}{\bibinfo{person}{Tobias Fiebig}, \bibinfo{person}{Martina Lindorfer}, {and} \bibinfo{person}{Seda G{\"u}rses}.} \bibinfo{year}{2022}\natexlab{}.
\newblock \showarticletitle{Position Paper: Escaping Academic Cloudification to Preserve Academic Freedom}.
\newblock \bibinfo{journal}{\emph{Privacy Studies Journal}} \bibinfo{volume}{1}, \bibinfo{number}{1} (\bibinfo{year}{2022}), \bibinfo{pages}{49--66}.
\newblock
\urldef\tempurl%
\url{https://doi.org/10.7146/psj.vi.132713}
\showURL{%
\tempurl}


\bibitem[\protect\citeauthoryear{Freitag, Berners-Lee, Widdicks, Knowles, Blair, and Friday}{Freitag et~al\mbox{.}}{2021}]%
        {freitag_real_2021}
\bibfield{author}{\bibinfo{person}{Charlotte Freitag}, \bibinfo{person}{Mike Berners-Lee}, \bibinfo{person}{Kelly Widdicks}, \bibinfo{person}{Bran Knowles}, \bibinfo{person}{Gordon~S. Blair}, {and} \bibinfo{person}{Adrian Friday}.} \bibinfo{year}{2021}\natexlab{}.
\newblock \showarticletitle{The real climate and transformative impact of {ICT}: {A} critique of estimates, trends, and regulations}.
\newblock \bibinfo{journal}{\emph{Patterns}} \bibinfo{volume}{2}, \bibinfo{number}{9} (\bibinfo{date}{Sept.} \bibinfo{year}{2021}), \bibinfo{pages}{100340}.
\newblock
\showISSN{26663899}
\urldef\tempurl%
\url{https://doi.org/10.1016/j.patter.2021.100340}
\showDOI{\tempurl}


\bibitem[\protect\citeauthoryear{Fu, Lin, and Lee}{Fu et~al\mbox{.}}{2014}]%
        {fu2014detecting}
\bibfield{author}{\bibinfo{person}{JuiHsi Fu}, \bibinfo{person}{PoChing Lin}, {and} \bibinfo{person}{SingLing Lee}.} \bibinfo{year}{2014}\natexlab{}.
\newblock \showarticletitle{Detecting spamming activities in a campus network using incremental learning}.
\newblock \bibinfo{journal}{\emph{Journal of network and computer applications}}  \bibinfo{volume}{43} (\bibinfo{year}{2014}), \bibinfo{pages}{56--65}.
\newblock
\urldef\tempurl%
\url{https://doi.org/10.1016/j.jnca.2014.03.010}
\showURL{%
\tempurl}


\bibitem[\protect\citeauthoryear{Kemp}{Kemp}{2025}]%
        {kemp_digital_2025}
\bibfield{author}{\bibinfo{person}{Simon Kemp}.} \bibinfo{year}{2025}\natexlab{}.
\newblock \bibinfo{title}{Digital 2025: email is still essential}.
\newblock
\newblock
\urldef\tempurl%
\url{https://datareportal.com/reports/digital-2025-sub-section-email-still-essential}
\showURL{%
\tempurl}


\bibitem[\protect\citeauthoryear{Lövehagen, Malmodin, Bergmark, and Matinfar}{Lövehagen et~al\mbox{.}}{2023}]%
        {lovehagen_assessing_2023}
\bibfield{author}{\bibinfo{person}{N. Lövehagen}, \bibinfo{person}{J. Malmodin}, \bibinfo{person}{P. Bergmark}, {and} \bibinfo{person}{S. Matinfar}.} \bibinfo{year}{2023}\natexlab{}.
\newblock \showarticletitle{Assessing embodied carbon emissions of communication user devices by combining approaches}.
\newblock   \bibinfo{volume}{183} (\bibinfo{year}{2023}), \bibinfo{pages}{113422}.
\newblock
\showISSN{1364-0321}
\urldef\tempurl%
\url{https://doi.org/10.1016/j.rser.2023.113422}
\showDOI{\tempurl}


\bibitem[\protect\citeauthoryear{Malmodin, L{\"o}vehagen, Bergmark, and Lund{\'e}n}{Malmodin et~al\mbox{.}}{2024}]%
        {malmodin2024ict}
\bibfield{author}{\bibinfo{person}{Jens Malmodin}, \bibinfo{person}{Nina L{\"o}vehagen}, \bibinfo{person}{Pernilla Bergmark}, {and} \bibinfo{person}{Dag Lund{\'e}n}.} \bibinfo{year}{2024}\natexlab{}.
\newblock \showarticletitle{ICT sector electricity consumption and greenhouse gas emissions--2020 outcome}.
\newblock \bibinfo{journal}{\emph{Telecommunications Policy}} \bibinfo{volume}{48}, \bibinfo{number}{3} (\bibinfo{year}{2024}), \bibinfo{pages}{102701}.
\newblock
\urldef\tempurl%
\url{https://doi.org/10.1016/j.telpol.2023.102701}
\showURL{%
\tempurl}


\bibitem[\protect\citeauthoryear{Pearce}{Pearce}{2020}]%
        {pearce2020energy}
\bibfield{author}{\bibinfo{person}{Joshua~M Pearce}.} \bibinfo{year}{2020}\natexlab{}.
\newblock \showarticletitle{Energy conservation with open source ad blockers}.
\newblock \bibinfo{journal}{\emph{Technologies}} \bibinfo{volume}{8}, \bibinfo{number}{2} (\bibinfo{year}{2020}), \bibinfo{pages}{18}.
\newblock
\urldef\tempurl%
\url{https://doi.org/10.3390/technologies8020018}
\showURL{%
\tempurl}


\bibitem[\protect\citeauthoryear{Pesari, Lagioia, and Paiano}{Pesari et~al\mbox{.}}{2023}]%
        {pesari2023client}
\bibfield{author}{\bibinfo{person}{Fabio Pesari}, \bibinfo{person}{Giovanni Lagioia}, {and} \bibinfo{person}{Annarita Paiano}.} \bibinfo{year}{2023}\natexlab{}.
\newblock \showarticletitle{Client-side energy and GHGs assessment of advertising and tracking in the news websites}.
\newblock \bibinfo{journal}{\emph{Journal of Industrial Ecology}} \bibinfo{volume}{27}, \bibinfo{number}{2} (\bibinfo{year}{2023}), \bibinfo{pages}{548--561}.
\newblock
\urldef\tempurl%
\url{https://doi.org/10.1111/jiec.13376}
\showURL{%
\tempurl}


\bibitem[\protect\citeauthoryear{Pärssinen, Kotila, Cuevas, Phansalkar, and Manner}{Pärssinen et~al\mbox{.}}{2018}]%
        {parssinen_environmental_2018}
\bibfield{author}{\bibinfo{person}{M. Pärssinen}, \bibinfo{person}{M. Kotila}, \bibinfo{person}{R. Cuevas}, \bibinfo{person}{A. Phansalkar}, {and} \bibinfo{person}{J. Manner}.} \bibinfo{year}{2018}\natexlab{}.
\newblock \showarticletitle{Environmental impact assessment of online advertising}.
\newblock \bibinfo{journal}{\emph{Environmental Impact Assessment Review}}  \bibinfo{volume}{73} (\bibinfo{date}{Nov.} \bibinfo{year}{2018}), \bibinfo{pages}{177--200}.
\newblock
\showISSN{01959255}
\urldef\tempurl%
\url{https://doi.org/10.1016/j.eiar.2018.08.004}
\showDOI{\tempurl}


\bibitem[\protect\citeauthoryear{Rao and Chien}{Rao and Chien}{2024}]%
        {rao_understanding_2024}
\bibfield{author}{\bibinfo{person}{Varsha Rao} {and} \bibinfo{person}{Andrew~A. Chien}.} \bibinfo{year}{2024}\natexlab{}.
\newblock \showarticletitle{Understanding the {Operational} {Carbon} {Footprint} of {Storage} Reliability and {Management}}. In \bibinfo{booktitle}{\emph{{HotCarbon}: {Workshop} on Sustainable {Computer} {Systems}}}.
\newblock
\urldef\tempurl%
\url{https://hotcarbon.org/assets/2024/pdf/hotcarbon24-final61.pdf}
\showURL{%
\tempurl}


\bibitem[\protect\citeauthoryear{{Statista}}{{Statista}}{2024a}]%
        {statistaemails2024}
\bibfield{author}{\bibinfo{person}{{Statista}}.} \bibinfo{year}{2024}\natexlab{a}.
\newblock \bibinfo{title}{Daily Number of Emails Worldwide from 2017 to 2025}.
\newblock
\newblock
\urldef\tempurl%
\url{https://www.statista.com/statistics/456500/daily-number-of-e-mails-worldwide/}
\showURL{%
\tempurl}
\newblock
\shownote{Accessed: 2024-10-02.}


\bibitem[\protect\citeauthoryear{{Statista}}{{Statista}}{2024b}]%
        {statistaspam2024}
\bibfield{author}{\bibinfo{person}{{Statista}}.} \bibinfo{year}{2024}\natexlab{b}.
\newblock \bibinfo{title}{Global spam volume as percentage of total e-mail traffic from 2011 to 2023}.
\newblock
\newblock
\urldef\tempurl%
\url{https://www.statista.com/statistics/420400/spam-email-traffic-share-annual/}
\showURL{%
\tempurl}
\newblock
\shownote{Accessed: 2024-10-02.}


\end{thebibliography}

\end{document}